\documentclass[preprint,showpacs,preprintnumbers,amsmath,amssymb]{revtex4}

\usepackage{graphicx}
\usepackage{dcolumn}
\usepackage{bm}

\begin{document}

\title{Leakage radiation microscopy of surface plasmon polaritons }

\author{A.~Drezet, A.~Hohenau, D.~Koller, A.~Stepanov, H.~Ditlbacher, B.~Steinberger, F.~R.~Aussenegg, A.~Leitner, and J.~R.~Krenn}
\affiliation{Institute of Physics, Karl-Franzens University Graz,
Universit\"atsplatz 5 A-8010 Graz, Austria}
\date{\today}

\begin{abstract}
We review the principle and methodology  of leakage radiation
microscopy (LRM) applied to surface plasmon polaritons (SPPs). Therefore we first analyse in detail the electromagnetic theory of leaky SPP waves. We
show that LRM is a versatile optical far-field method allowing
direct quantitative imaging and analysis of SPP propagation on
thin metal films.  We illustrate the LRM potentiality by
analyzing the propagation of SPP waves interacting with several two dimensional
plasmonic devices realized and studied in the recent
years.\end{abstract}

 \maketitle
\section{Introduction}

In recent years intensive investigations of surface plasmon
polaritons (SPPs) have been made in the promising context of
nanophotonics. This research is actually motivated by the
current trends for optical device miniaturisation and by the
possibilities of merging aspects of nanophotonics with those of
electronics. SPPs are electromagnetic waves bounded to
dielectric-metal interface. As surface waves, SPPs are
exponentially damped in the directions perpendicular to the
interface \cite{Raether}. Furthermore, SPPs could be used to
transfer optical information in a two dimensional (2D) environment.
This appealing property can be used for optical addressing of
different 2D optical systems and nanostructures located at a
dielectric/metal interface. Actually several 2D  SPP devices
including passive nanostructures including mirrors or beam splitter
and active elements like molecules or quantum dots are currently
under development and investigation. Developments such as these
raise the prospect of a new branch of photonics using SPPs, for
which the term "plasmonics" emerged \cite{Barnes:2003,Ebbesen:2007,Drezet:Micron}. \\

However, for experimental investigations  of optical devices an
important characteristic of SPP modes is that their spatial
extent is governed and defined by the geometry of the nanoelements
rather that by the optical wavelength \cite{Krenn:1999}. This
consequently opens possibilities for breaking the diffraction
limit but requires instruments of observation adapted essentially
to the subwalength regime and being capable of imaging the
propagation of SPPs in their 2D environment. Usually the analysis of
the subwalength regime implies necessarily near field optical (NFO)
methods \cite{Pohl,Courjon:2003} able to collect the evanescent (i.~e., non
radiative) components of the electromagnetic fields associated
with SPPs. However, when the metal film on which the 2D optical
elements are built is thin enough (i.~e~., with a thickness below
80-100 nm) and when the subtratum optical constant (usually glass)
is higher than the one of the superstratum medium an other
possibility for analyzing SPP propagation occurs. This possibility
is based on the detection of coherent leaking of SPPs through the
substratum. Such a far-field optical method is called leakage
radiation microscopy (LRM) \cite{Hecht,Bouhelier,Stepanov} and allows indeed a direct quantitative
imaging and analysis of SPP propagation on thin metal films.\\
The aim of this article is to present a short overview of recent
progress in the field of SPP imaging using LRM. In a first part of
this work we will describe the theoretical principles underlying
LRM. In the second part we will discuss modern leakage radiation
methods and illustrate the LRM potentialities by analyzing few
experiments with SPP waves interacting with 2D plasmonic devices.
\section{Leakage radiation and surface plasmon polaritons}
In order to describe the theoretical mechanisms explaining leakage
radiation it will be sufficient for the present purpose to limit
our analysis to the case of a metal film of complex permittivity
$\epsilon_1(\omega)=\epsilon'_1+i \epsilon''_1$ ($\omega=2\pi
c/\lambda$ is the pulsation) sandwiched between two dielectric
media of permittivity $\epsilon_0$ (substrate) and
$\epsilon_2<\epsilon_0$ (superstratum). This system is
theoretically simple and to a good extent experimentally
accessible \cite{Raether,Burke}. In the limiting case where the
film thickness $D$ is much bigger than the SPP penetration length
in the metal (i.~e., $D\gtrsim 70$ nm for gold or silver in the
visible domain) one can treat the problem as two uncoupled single
interfaces. We will consider as an example the interface 0/1 (the
media 0 and 1 are located in the domain $z\geq 0$ and $z\leq 0$
respectively ). Such an interface will be identified in the
following with the plane $z=const.$ in cartesian coordinates. An
elementary harmonic SPP wave is actually a TM electromagnetic mode
characterized by its pulsation $\omega$ and its magnetic field
$\mathbf{H}=[0,H_y,0]$ where the y component can be written
\begin{eqnarray}
H_0=\alpha e^{ik_x x}e^{ik_{z0} z}e^{-i\omega t} & \textrm{in the medium 0} \nonumber \\
H_1=e^{ik_x x}e^{ik_{z1} z}e^{-i\omega t} & \textrm{in the medium
1,}
\end{eqnarray}and where $k_{x}=k'_x+ik''_x$ is the (complex valued) wavevector of the SPP propagating in the x direction along the interface.
$k_{zj}\equiv k_{j}=\pm\sqrt{[(\omega/c)^2\epsilon_i-k_x^2]}$ are
the wave vectors in the medium j =[0 (dielectric), 1 (metal)]
along the direction $z$ normal to the interface. By applying
boundary conditions to Maxwell's equations one deduces
additionally $\alpha=1$ and
\begin{eqnarray}
 \frac{k_1}{\epsilon_1}-\frac{k_0}{\epsilon_0}=0,
\end{eqnarray}
which implies
\begin{equation}
k_{x}=\pm(\omega/c)\sqrt{\frac{\epsilon_{0}\epsilon_1}{\epsilon_{0}+\epsilon_1}}
\end{equation}
\begin{equation}
k_{j}=\pm(\omega/c)\sqrt{\frac{\epsilon_{j}^2}{\epsilon_{0}+\epsilon_1}}
\end{equation}
for a SPP wave propagating along the x direction. The choice of
the sign convention connecting the z and x  components of the wave
vector is a priori arbitrary and must be done only on a physical
ground.\begin{figure}[hbtp]
\includegraphics[width=8cm]{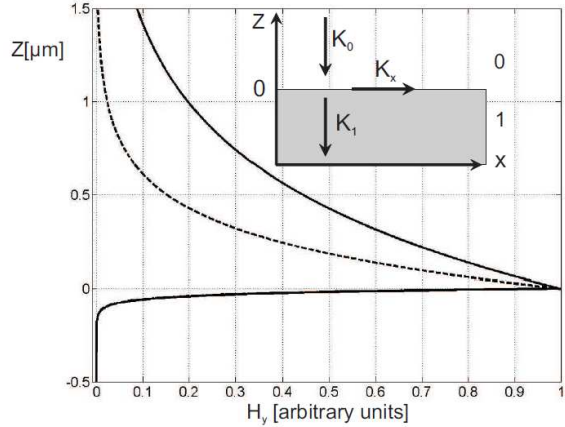}
\caption{Structure of the SPP magnetic field $\textrm{Real}[H_y]$
across an interface air/gold (thick line) and glass/gold (dashed
line). The optical wavelength considered is $\lambda=800$ nm. The
permittivity of glass is taken to be
$\epsilon_{\textrm{glass}}=2.25$. The inset shows the conventions
for the axes x and z. The arrows indicate the direction of the
real part of the wave vector normally and parallely to the
interface.}
\end{figure} Indeed, due to ohmic losses in the metal we expect an exponentially decaying SPP wave
propagating along the interface. This condition implies the
relation $k'_x\cdot k''_x\geq 0$ \cite{Burke}. This inequality is
actually always fulfilled since from Eq.~3 one deduces
\begin{equation}
k'_x\cdot
k''_x=\frac{1}{2}(\omega/c)^2\frac{\epsilon_{0}^2\epsilon''_1}{(\epsilon_{0}+\epsilon'_1)^2+(\epsilon''_1)^2}>
0
\end{equation}
which is indeed positive because $\epsilon''_1>0$. By writing
$k_{zj}=k'_j+ik''_j$ one additionally obtains the relation
\begin{equation}
-k'_0\cdot k''_0=(\omega/c)^2\epsilon''_{1}/2-k'_1\cdot
k''_1=k'_x\cdot k''_x\geq 0.
\end{equation}
This relation fixes the sign conventions since the wave must also
decay exponentially when going away from the interface in both
media. More precisely one gets
\begin{equation}
k'_0\cdot k''_0\leq 0,
\end{equation}
\begin{equation}
k'_1\cdot k''_1=\geq 0. \end{equation}The product $k'_1\cdot
k''_1$ is  positive if $\epsilon'_1\geq
-|\epsilon_1|^2/(2\epsilon_0)$, a fact which is indeed true for
silver and gold interfaces with air or glass in most of the
visible optical domain. However small negative values of Eq.~8
occur for silver close to the interband region around $\lambda\sim
350 nm$. Additionally a higher value of $\epsilon_0$ will also
change the sign in Eq.~8. Fig.~1 shows the behavior of the SPP
magnetic field close to an interface gold/air and gold/glass at
the optical wavelength $\lambda=800$ nm. At such a wavelength the
conditions given by Eqs.~5-8 impose the solutions
\begin{eqnarray}
k_{x}=\pm(\omega/c)\sqrt{\frac{\epsilon_{0}\epsilon_1}{\epsilon_{0}+\epsilon_1}},&
k_{j}=-(\omega/c)\sqrt{\frac{\epsilon_{j}^2}{\epsilon_{0}+\epsilon_1}}.
\end{eqnarray}
The real parts of the $k_z$ components of the SPP wave vector are
for both media oriented in the same direction corresponding  to a
wave propagating from the air side to the metal side (see inset in
Fig.~1). Furthermore the waves are exponentially damped when going
away from the interface  in agreement with Eq.~7, 8 (see Fig.~1).
Most important for us is that the Poynting vector \cite{Jackson}
$\mathbf{S}=\textrm{Real}[\mathbf{E}\times \mathbf{H}^{\ast}]/2$
is defined in the medium j by
\begin{eqnarray}
\mathbf{S}_j=\frac{1}{2}c\textrm{Real}[\frac{k_x
\hat{\mathbf{x}}+k_j\hat{\mathbf{z}}}{\omega\epsilon_j/c}]e^{-2k''_x
x-2k''_z z}.
\end{eqnarray}
On the dielectric side the energy flow is as expected oriented in
the direction of $\textrm{Real}[\mathbf{k}]$. However it can be
shown on the metal side and for wavelengths not too close from the
spectral region associated with the interband transition of gold
or silver  that the energy flow in the x direction is oriented
oppositely to the wave vector $\textrm{Real}[k_x]$ since
$\textrm{Real}[k_x/\epsilon_1]=(k'_x\epsilon'_1+k''_x\epsilon''_1)/|\epsilon_1|^{2}$
is dominated by $k'_x\cdot \epsilon'_1$ and since $\epsilon'_1<0$.
However the total energy flow  in the x direction $S_x=
\int_0^{+\infty}S_x^{(\textrm{air})}dz+\int_{-\infty}^0
S_x^{(\textrm{metal})}dz$ is oriented along $k'_x$. Additionally
in the z direction
$\textrm{Real}[k_1/\epsilon_1]=(k'_1\epsilon'_1+k''_1\epsilon''_1)/|\epsilon_1|^{2}$
is parallel to $k'_1$ since $k''_1\epsilon''_1$ dominate. This
implies in particular that the energy associated with the SPP wave
is absorbed by the metal during its propagation through the
interface from the air side to the glass side. It should be
observed that already in the case of the ideal plasma model
neglecting losses the wave vector $k_x=k'_x$ is antiparallel to
$S_x$ in the metal since there is no imaginary part and since
$\epsilon_1=1-\omega_p^2/\omega^2<0$
(see also \cite{Zhinzhin,Kats}).   \\
We show in Fig.~2 the curves associated with the dispersion
relations of SPPs propagating along a gold/air and gold/glass
interface respectively. Fig.~2A represents the dependencies
$\omega$ versus $k'_x$ corresponding to Eq.~3. Fig.~2B shows the
dependencies $\omega$ versus $L_{SPP}$ where $L_{SPP}=1/(2k''_x)$
is the propagation length of the SPP waves (for the metal optical
constant we used the experimental values given in \cite{Johnson}).
The typical back bending around $\lambda=520 nm$  corresponds to
the resonance associated with the bound electrons and the
interband transition (for a good discussion see
\cite{Novotny:2006}). Far away from the interband the real part of
the dispersion is close to the asymptotic light lines: we speak
about Zenneck surface modes by opposition to Fano and evanescent
modes existing close to the interband \cite{Zhinzhin,Halevi}.
\begin{figure}[hbtp]
\includegraphics[width=8cm]{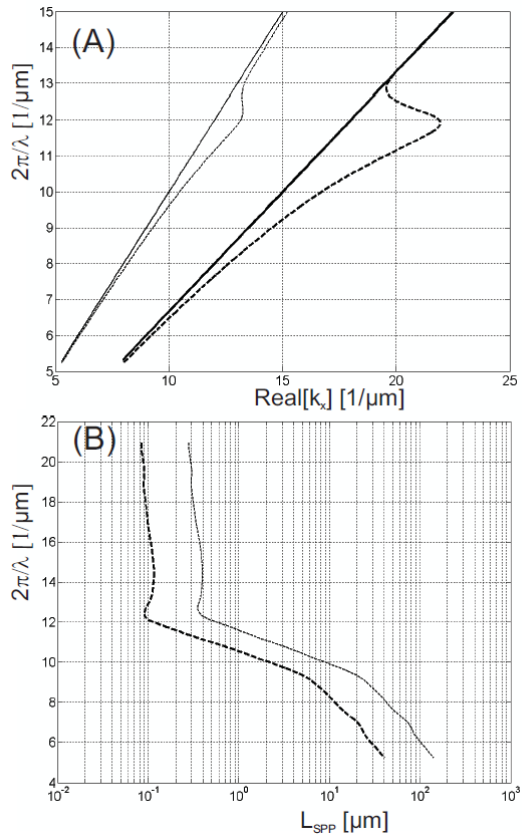}
\caption{Dispersion relations of SPP waves propagating along a
air/gold (thin dashed curve) and glass/gold (thick dashed curve)
interface. (A) Real part of the dispersion relations showing the
evolution of $k'_x$ with $\omega/c=2\pi/\lambda$. The light cones
corresponding to an optical wave propagating in air and glass
(i.~e., in the bulk medium) are represented by a thin and thick
continuous line respectively. The back bending at $\lambda=520$ nm
correspond to the interband resonance. (B) Imaginary part of the
dispersion relations showing the evolution of $L_{SPP}$ with
$\omega/c=2\pi/\lambda$. The damping is higher for the glass/gold
interface.}
\end{figure}
An important feature occurs close to this interband transition
since the slope in Fig.~2A is diverging. This means that the group
velocity defined by $v_g=\partial\omega/\partial k'_x$ is
infinite. One can qualitatively deduce that there is however no paradox with causality by remarking that the
propagation length is approaching zero at such wavelengths.
Actually the  SPP life time $\tau_{SPP}$ defined by $v_g \cdot
\tau_{SPP}\approx L_{SPP}$ is tending as well to zero. This is a
clear signature of the absence of significant SPP propagation in the
interband spectral region. This fact actually precludes faster than light information
transfer by SPP waves and is in agreement with relativistic causality.\\
A second important feature concerning Fig.~2 A is that the
air/gold dispersion curve is located inside of the light cone for
glass defined by the equation
$\mathbf{K}^2=(\omega/c)^2\epsilon_{\textrm{glass}}$ where
$\mathbf{K}$ is a real light wave vector and
$\epsilon_{\textrm{glass}}=2.25$. Writing $\mathbf{K}=[K_x,0,K_z]$
the wavevector of a TM (i.~e., p polarized) plane wave propagating
away from the interfaces into the glass side one see that SPPs
propagating at the air/metal interface can radiate into the glass
substrate if the condition
\begin{equation}
K_{x}\approx\pm
\textrm{Real}[(\omega/c)\sqrt{\frac{\epsilon_{0}\epsilon_1}{\epsilon_{0}+\epsilon_1}}]
\end{equation} is approximately fulfilled.
Here we neglected the role of the imaginary part in Eq.~3. Similarly one can deduce that none of the SPPs propagating at the
two interfaces can radiates into the air side. \\
In order to have a more complete analysis one must actually consider the
problem with two coupled interface 0/1 (glass/metal) and 1/2 (metal/air) supporting SPP
waves and separated by a small distance $D$. The two interfaces are coupled by evanescent SPP waves tunnelling through the metal slab.
Such a mathematical problem can only be treated numerically by resolving an implicit equation. As
for the single interface this equation can be defined
directly from Maxwell Equations \cite{Burke}. However it is much easier and convenient for the following
to remark  with Raether \cite{Raether} that Eq.~2 and consequently Eq.~3 are
obtained by finding the zeros of the numerator in the Fresnel
reflectivity coefficient for a TM wave coming from the dielectric
side:
\begin{equation}
R_{0,1}^{p}=\frac{(k_{0}/\epsilon_{0}-k_{1}/\epsilon_{1})}{(k_{1}/
\epsilon_{1}+k_{0}/\epsilon_{0})},\end{equation} with
$k_{j}=\pm\sqrt{((\omega/c)^2\epsilon_{j}-k_{x}^{2})}$. Actually
Raether \cite{Raether} reasoned with the denominator of the
Fresnel coefficient due to different conventions for the signs of
the wave vectors $k_j$. However it is remarkable that the result
is the same at the end of the calculations. Identically one can
thus define the Fresnel coefficient for a TM wave reflected by the
slab 0-1-2 \cite{Jackson,Novotny:2006,Raether}:
\begin{equation}
R_{0,1,2}^{p}=\frac{R_{0,1}^{p}+R_{1,2}^{p}e^{2ik_{1}D}}{1+R_{0,1}^{p}R_{1,2}^{p}e^{2ik_{1}D}},\end{equation}
and find the zeros of the numerator, i.~e., one can solve the
implicit equation
\begin{equation}
 R_{0,1}^{p}+R_{1,2}^{p}e^{2ik_{1}D}=0,
\end{equation}
in order to define the SPP dispersion relations. From this
equation it follows that
\begin{eqnarray}
(k_{0}/\epsilon_{0}+k_{1}/\epsilon_{1})(k_{2}/\epsilon_{2}-k_{1}/\epsilon_{1})e^{ik_{1}D}\nonumber\\
+(k_{1}/\epsilon_{1}-k_{0}/\epsilon_{0})(k_{2}/\epsilon_{2}+k_{1}/\epsilon_{1})e^{-ik_{1}D}=0.
\end{eqnarray}
As for the single interface one has an important relation between
the real and imaginary parts of the SPP wave vectors in the
different medium:
\begin{equation}
-k'_0\cdot k''_0=(\omega/c)^2\epsilon''_{1}/2-k'_1\cdot
k''_1=-k'_2\cdot k''=k'_x\cdot k''_x.
\end{equation}
Since we are interested only into the solutions which are decaying
along the interface we (in agreement with our previous treatment
of the single interface) suppose the condition $k'_x\cdot
k''_x\geq 0$ satisfied. A second important point is that due to
the arbitrariness in the sign of $k_j$ there are in fact apriori 8
possibilities for writing Eq.~15.  However, Eq.~15 is invariant
under the transformation $k_1 \rightarrow -k_1$. This means that
the number of apriori possibilities for the sign of $k_j$ is
reduced from 8 to 4. This multiplicity was studied by Burke
\emph{et al.} \cite{Burke} however since for the present purpose
we are looking for SPP waves leaking from the air/metal interface
into the glass substrate we will consider only the possibility
\begin{eqnarray}
k_{0}=-\sqrt{((\omega/c)^2\epsilon_{0}-k_{x}^{2})}\nonumber\\
k_{1}=+\sqrt{((\omega/c)^2\epsilon_{1}-k_{x}^{2})}\nonumber\\
k_{2}=-\sqrt{((\omega/c)^2\epsilon_{2}-k_{x}^{2})}.
\end{eqnarray}
The sign of $k_1$ is however arbitrary as explained above and we
choose it here positive by definition. In order to define a SPP
wave leaking into the glass substrate one has thus to substitute
Eq.~17 into Eq.15 and find numerically (i.~e., by a minimization
procedure \cite{Hohenau,Dionne}) the zeros of the implicit
equation with variable $k'_x$ and $k''_x$. \begin{figure}[h]
\includegraphics[width=8cm]{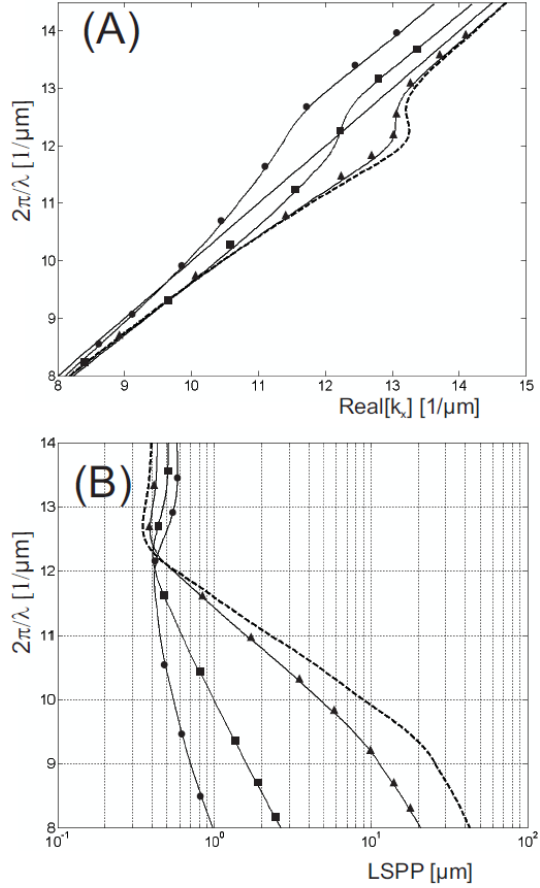}
\caption{Dispersion relations of SPP waves propagating along an
air/gold interface and leaking into glass. The dispersion is
calculated  for four  film thickness $D$ : 70 nm (dashed line
curve)es, 50 nm (continuous line with triangular markers), 20 nm
(continuous line with square markers), and 10 nm (continuous line
with circular markers).(A) Real part of the dispersion relations
showing the evolution of $k'_x$ with $\omega/c=2\pi/\lambda$. The
light cone corresponding to optical wave propagating in air
(i.~e., in the bulk medium) is represented by a continuous line.
(B) Imaginary part of the dispersion relations showing the
evolution of $L_{SPP}$ with $\omega/c=2\pi/\lambda$.}
\end{figure}This has to be done only in the
quarter of the complex plane corresponding to $k'_x>0$, $k_x''>0$.
The quarter $k'_x<0$, $k_x''<0$ must  be equivalent due to
symmetry and corresponds actually to decaying SPP waves
propagating in the negative x direction. The two other quarters of
the complex plane correspond to growing SPP waves along the
interface and will be rejected on a physical ground (compare
\cite{Burke}).

Figs.~3A and 3B show numerical calculations of dispersion
relations corresponding to a SPP wave leaking through a gold film
from the air side to the glass side.  The thickness is taken to be
$D$=70, 50, 20, and 10 nm respectively. For the value $D\geq 70$
nm the dispersion relation is identical to the dispersion for the
single air/gold interface for semi infinite media. However, for
smaller thickness the coupling between the interface increases and
the propagation length decreases as shown on Fig.~3 B.
\begin{figure}[hbtp]
\includegraphics[width=8cm]{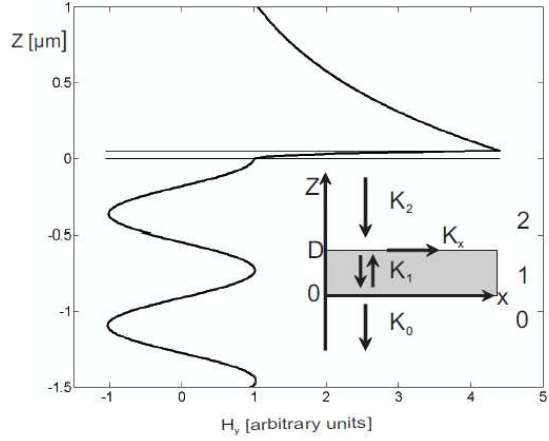}
\caption{ Structure of the magnetic field $\textrm{Real}[H_y]$
associated with a SPP mode leaking through a gold film of
thickness 50 nm (medium 1) from the air side (medium 2) to the
glass side (medium 0). The evolution is represented along the
normal $z$ to the interfaces. The optical wavelength considered is
$\lambda=800$ nm. The two horizontal lines show the interfaces
separated by 50 nm. The inset shows the conventions for the axis x
and z. The arrows indicate the direction of the real part of the
wave vector normally and parallely to the interface.}
\end{figure}
The magnetic field associated with SPP electromagnetic modes in this layered system is given by
\begin{eqnarray}
H_0=e^{ik_x x}e^{ik_{0} z}e^{-i\omega t} \nonumber \\
H_1=e^{ik_x x}[\alpha\sin{(k_{1}z)}+\beta\cos{(k_{1}z)}]e^{-i\omega t}\nonumber \\
H_2=\gamma e^{ik_x x}e^{ik_{2} (z-D)}e^{-i\omega t}.
\end{eqnarray}
The coefficients $\alpha, \beta, \gamma$ are obtained by
considering the boundary conditions and one finds
\begin{eqnarray}
\alpha=i\frac{k_0\epsilon_1}{k_1\epsilon_0}, \beta=1 \nonumber\\
\gamma=i\frac{k_0\epsilon_1}{k_1\epsilon_0}\sin{(k_{1}D)}+\cos{(k_{1}D)}.
\end{eqnarray}
As an illustration we show in Fig.~4 the evolution of the real
part of the magnetic field across a 50 nm thick gold slab
sandwiched between the glass substratum and the air superstratum
for an optical wavelength $\lambda=800 nm$. As visible the SPP
field is located in the vicinity of the air/gold interface and  is
evanescent on the air side. This is clearly reminiscent of our
former analysis of SPPs propagating at the single air/metal. In
addition however the wave is leaking radiatively (i.~e.,
propagatively) into the glass substrate. However from Eq.~18 and
the condition $k'_0\cdot k''_0\leq 0$ it is clear that the leaking
wave is exponentially growing in the -z direction when going away
from the gold slab. This is already the result we obtained when we
considered the limit of the thick slab. An exponentially growing
wave looks non physical and is in particular associated with
infinite radiated energy in the far field. There is now the
question of how to connect a growing wave with the basic reasoning
giving rise to Eq.~11 and the idea of phase matching between the
(real part) of the SPP wave vector with a propagative plane wave
vector in the glass substrate. However such paradoxes disappear
for two reasons: First, an infinite energy occurs only because we
considered an infinite interface or equivalently because we did
not consider how the SPP is locally launched on the metal film.
When such conditions are taken into account this paradox must
disappear \cite{Burke}. Second, the SPP wave defined by Eq.~18 is
actually a wave packet when looked at through the Fourier basis of
propagative TM plane waves. Since in the far field (i.~e., in the
glass substrate) one actually detect such plane waves one must do
a Fourier transform in order to generalize Eq.~11
\cite{Raether,Burke}. Instead of Eq.~11 one obtains consequently a
statistical distribution of (real) wavevectors $K_x$ given by
\begin{equation}
I(K_x)=\frac{\textrm{const.}}{(K_x-k'_x )^2+(k_x'')^2},
\end{equation}
where $2k_x''$ defines the full width at half maximum (FWHM) of
this Lorentz distribution of radiated plane waves. By noting as
usual $\theta$ the angle between the wave vectors $\mathbf{K}$ of
the plane waves refracted into the glass substrate and the normal
$z$ to the interfaces one has by definition
$K_x=2\pi\sqrt{\epsilon_0}/\lambda\sin{\theta}$ and the angular
distribution of radiated power is in the far field given by:
\begin{equation}
I(\theta)=\frac{\textrm{const.}}{(2\pi\sqrt{\epsilon_0}/\lambda\sin{\theta}-k'_x )^2+(k_x'')^2}.
\end{equation}
\section{Leakage radiation microscopy}
Historically the first observations of leakage radiation by SPP
propagating on a thin metal film were reported by analyzing
scattering by rough metal surfaces of light into SPPs
\cite{Raether,Simon}. The possibility of using rough surface to
excite SPPs was extensively studied in the past \cite{Raether} and
is based on the fact that the scattering by small defects on a
flat film can represent a source of evanescent momentum sufficient
for the light waves to match the SPP dispersion relation.
Equivalently the amount of momentum needed can be carried by
grating coupling \cite{Raether}. SPP waves are subsequently
emitted back into the glass substrate as leakage radiation (see
Fig.~5).
\begin{figure}[hbtp]
\includegraphics[width=6cm]{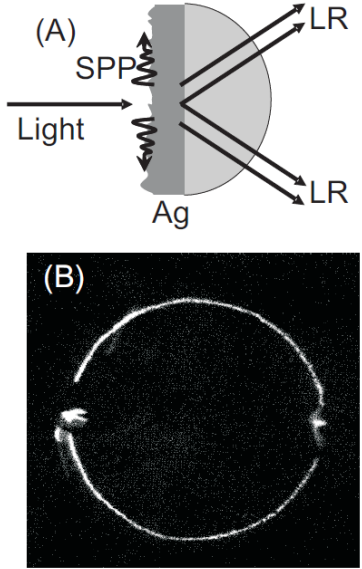}
\caption{Observation of leakage radiation (LR) through a rough
thin silver film (wavelength $\lambda=550 nm$). (A) SPPs are
launched on the air/silver interface by using scattering by film
rugosities to produce the amount of momentum necessary to match
the SPP dispersion relation. SPPs are leaking through the glass
substrate and detected using a photographic plate in the
far-field. (B) Photography showing the leakage radiation cone. The
photography is extracted from the work by H.~J Simon and
J.~K.~Guha \cite{Simon} ($\copyright$ Opt.~Comm., Elsevier, 1976).
}
\end{figure}
This light collected on a photographic plate forms a ring-like
distribution in agreement with Eqs.~11, 20, 21. The FWHM of the
SPP wavevector distribution is in direct correspondence with the
radial width of the ring \cite{Simon}. Further progress was
obtained recently with the development of near field scanning
optical microscopy (NSOM) which allows the local optical
excitation of evanescent waves in the vicinity of a metal surface.
Such evanescent waves can carry a sufficient amount of momentum to
couple to SPP waves. Direct observations have indeed confirmed
this principle \cite{Sonnichsen,Brun:2003,Hecht}. As an example we
show in Fig.~6 an experiment in which the NSOM tip launches SPPs
on an aluminum film which after interaction with a hole excites
optically some  quantum dots (QDs) located below. The collected
signal shows a specific QD luminescence spectrum \cite{Brun:2002}.
By scanning the sample around the NSOM tip one can realize SPP
mapping since the hole acts a probe structure for the field
emitted by the tip. Quantitative analysis of the total
luninescence of the QDs associated with a given hole show clearly
that the QD excitation is mediated by SPPs propagating on the
aluminum film.
\begin{figure}[hbtp]
\includegraphics[width=6cm]{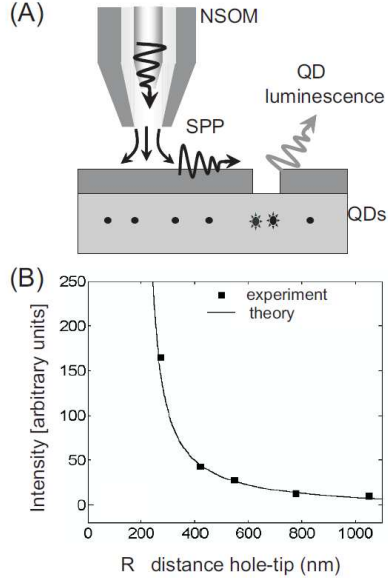}
\caption{(A) Sketch of the experiment using a NSOM tip to excite
SPPs on an aluminum film with nanoholes. The film containing  240
nm diameter holes covers a CdTe/ZnTe system of QDs whose spectral
luminescence induced by SPP excitation is characteristic of a
given hole. The laser excitation wavelength is 514.5 nm. (B)
Radial mapping of the SPP intensity. The SPP intensity is
proportional to the QDs luminescence of a given hole. The curve
represents the evolution of the total luminescence collected
(using a detector in the optical far field) as a function of the
distance $R$ separating the NSOM (source) from the hole (probe).
The experiment is made at 4.2 K \cite{Brun:2003} ($\copyright$
Europhys.~Lett., EDP, 2003).}
\end{figure}
Fig.~6B shows the radial dependence of the collected intensity.
These results agree well with a 2D SPP dipole model supposing an
effective dipole located at the tip apex \cite{Hecht,Brun:2003}
(see also the discussions concerning the Bethe-Bouwkamp
\cite{Bethe,Bouwkamp1,Bouwkamp2} theory of diffraction by a small
aperture in a metal film in
\cite{Karrai1,Karrai2,Drezet:EPL2001,Drezet:PRE2002,Drezet:EPL2004}).
Following this model the SPP wave can be theoretically modelled by
a scalar wave $\Psi(\rho,\theta)$ given by
\begin{equation}
\Psi(\rho,\phi)=\textrm{const.} \frac{e^{ik_{\textrm{SPP}}
\rho}\cos{\phi}}{\sqrt{\rho}},
\end{equation}
where $k_{\textrm{SPP}}=k'_x+ik''_x$ is given by Eq.~3, $\rho,
\phi $ are polar coordinates on the metal film and the origin of
the coordinate is taken at the dipole position. $\phi$ is the
angle between the dipole associated with the NSOM tip (parallel to
the polarization of the laser beam injected in the NSOM tip) and
the the radial vector $\mathbf{\rho}=[x,y]$. This simplified model
can be theoretically justified by using the Green Dyadic Formalism
\cite{Bozhe} and has been applied by many authors successfully
\cite{Hecht,Bouhelier,Brun:2003,Harry1,Harry2,Drezet:APL2005,Laluet}
to SPP waves propagating in various environments.\\
Several authors applied NSOM methods coupled to LRM
\cite{Hecht,Bouhelier}. In particular in \cite{Hecht} Hecht
\emph{et al.} realized an optical setup using an immersion oil
objective to collect the leakage radiation emitted by the NSOM tip
on gold or silver films (see fig.~7).
\begin{figure}[hbtp]
\includegraphics[width=6cm]{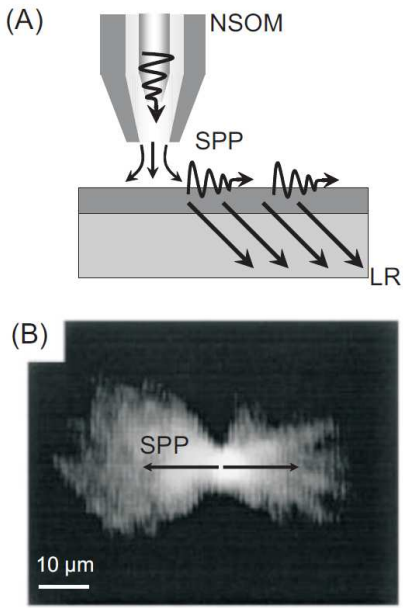}
\caption{(A) Principle of the experiment to generate SPP leakage
radiation (LR) using a NSOM tip. LR is collected using an
immersion oil objective (not shown here). (B) Map of SPP intensity
using a charge coupled device (CCD) camera to collect the LR
emitted through the glass substrate. SPPs are launched on a 60 nm
thick silver film at the optical laser wavelength $\lambda=633nm$.
The image in B is taken from B.~Hecht \emph{et al.} \cite{Hecht}
($\copyright$ Phys.~Rev.~Lett, American Physical Society, 1996). }
\end{figure}
The system shown in Fig.~7 B is a 60 nm thick silver film
optically excited by a NSOM tip at the laser wavelength
$\lambda=633nm$. It can be shown by analyzing Fig.~7 B that the
radiation pattern is well described by a 2D dipole model in
agreement with Eq.~22. In particular the SPP propagation length
was measured and is in fair agreement with our analysis in
Sect.~1. Additionally it was shown in \cite{Hecht} that one can
also analyze the Fourier distribution of SPP momentum (given by
Eqs.~20, 21)
by defocusing the objective lens. As expected SPP rings similar to the one of Fig.~5 were observed.\\
\begin{figure}[hbtp]
\includegraphics[width=6cm]{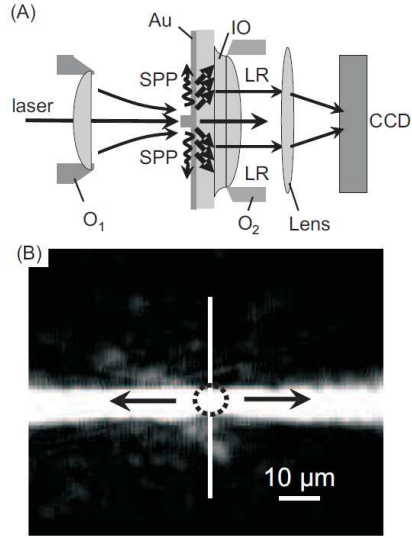}
\caption{(A) Principle of the experiment to generate SPP leakage
radiation (LR) using a focussed laser beam (objective $O_1$, 10x,
numerical aperture $NA=0.3$). LR is collected using an immersion
oil objective (objective $O_2$, 63x, $NA=1.3$) and refocussed on
the CCD camera by using an auxiliary lens. (B) Map of SPP
intensity using a CCD camera to collect the LR emitted through the
glass substrate. SPPs are launched on a 50 nm gold film at the
optical laser wavelength $\lambda=800nm$ from a gold ridge (50 nm
height, 150 nm width) represented by a white line. The size of the
laser spot with  a diameter $\simeq 8 $ $\mu$m is represented by
the white dashed circle.}
\end{figure}
In the same context  we developed in recent years a systematic
approach using far-field microscopy to analyze quantitatively the
interaction between SPPs and plasmonic devices by using LRM. The
nanodevices studied were all fabricated by electron beam
lithography (EBL) allowing the precise and reproducible tailoring
of metal and dielectric surfaces on a lateral size dimension down
to 20 nm \cite{EBL}. As an example we show in Fig.~8 B a LRM image
obtained by using a gold ridge (50 nm height, 150 nm width)
lithographed on a 50 nm thick gold film to launch two well
collimated and counter propagating SPP beams. These beams are
launched by focussing a laser beam with a microscope objective
(10x, numerical aperture $NA=0.3$) on the gold ridge. Scattering
by the nanostructure gives rise to evanescent waves supplying the
right amount of momentum necessary for generating a SPP wave. The
optical LRM setup is sketched in Fig.~8 A. Leakage radiation
emitted through the glass substrate is collected by an immersion
oil objective (63x, numerical aperture, $NA=1.3$). Light is
subsequently refocussed on a charge coupled device (CCD) camera.
The direct mapping of the SPP intensity with this method provides
a one-to-one correspondence between the 2D SPP intensity and the
image recorded on the CCD camera. It should be observed that an
incident laser beam with diameter $Dx$ is in the focal plane
(object plane) of the objective $O_1$ focussed into a disc of
diameter $w$ (i.~e., $w=$beam waist) such that
\begin{equation}
\tan{\alpha}=\frac{2\lambda}{\pi w}=\frac{Dx}{2f}
\end{equation}
where $f$ is the focal length of the objective  and $\alpha$ the
divergence angle of the laser beam focussed on the sample. The
direct application of this version of the Heisenberg relation
\cite{Teich} implies that the divergence angle $\alpha_{SPP}$ of
the SPP beam launched on the metal film must equal the divergence
angle $\alpha$ of the impinging laser beam. This result is in good
agreement with the experimental case shown in Fig.~8 with $Dx=2$
mm, $w=8.3$ $\mu$m and $\alpha=3.5^{\circ}$. Changing the
objective focal length is a straightforward means to
obtain different divergence angles $\alpha_{SPP}$ (see for example \cite{Drezet:APL2006,Stepanov,Drezet:EPL2006}).\\
\begin{figure}[h]
\includegraphics[width=8cm]{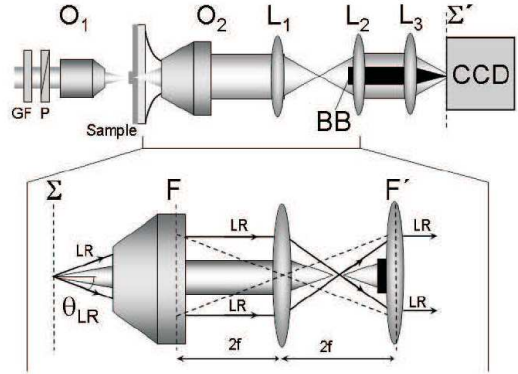}
\caption{Experimental scheme for LRM. SPPs are excited by laser
light focused by the microscope objective O1 (50x, numerical
aperture, $NA=0.7$) onto a structured gold thin film on a glass
substrate. LR emitted into the glass substrate from the gold/glass
interface $\Sigma$ is collected by the immersion microscopy
objective O2 (63x, NA=1.25) and imaged by a CCD camera. Depending
on the lateral position of the convex lens L3 either the back
focal plane or the sample plane is imaged. BB beam block, L1,2,3
convex lenses, f focal length of L1 and L2; $\Sigma$, $\Sigma'$,
sample plane and image thereof; F, F' back focal plane and image
thereof \cite{Drezet:APL2006}.}
\end{figure}
As a further improvement it is possible to modify the previous
optical setup in order to image not only the direct space
information but also the momentum corresponding to the Fourier
space. It is indeed a well known fact of Fourier optics
\cite{Teich,Born,Novotny:2006} that such a mapping of the wave
vector distribution (as given by Eq.~20) is in principle always
possible by recording the LR light in the back focal plane $F$ of
the oil immersion objective. In the optical setup shown in Fig.~9
\cite{Drezet:APL2006} we realized a dual microscope able to image
SPP propagation in both the direct and Fourier space. In
particular the back focal plane $F$ of the oil immersion objective
$O_2$
\begin{figure}[h]
\includegraphics[width=8cm]{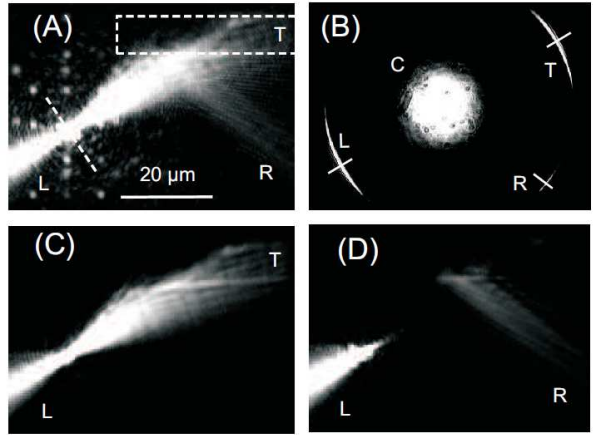}
\caption{(A) Direct space LRM imaging of two SPPs beams launched
from the ridge and propagating to the lower left ($L$) and upper
right to be partly reflected ($R$) and transmitted ($T$) by the
Bragg mirror. The image is taken with no beam block. Ridge and
Bragg mirror are indicated by the dashed line and rectangle,
respectively. (B) Fourier plane LRM imaging corresponding to (A).
C indicates the wave vector components of the directly transmitted
laser beam $C$; $R,L,T$ have the same meaning as in (A). (C)-(D)
are Fourier filtered images after removing the central beam $C$
with a beam block in $F'$ (see text). In (C) the reflected beam R
is removed by blocking the arc of the SPP ring \emph{R} shown in
(B). In (D) the transmitted beam T and the beam incident on the
Bragg mirror are removed by blocking the arc of SPP ring $T$ in
(B). Data from \cite{Drezet:APL2006} ($\copyright$,
Appl.~Phys.~Lett., American Institute of physics, 2006).}
\end{figure}
imaged onto a CCD camera in Fig.~9. With such a microscope it is
furthermore possible to act experimentally in the Fourier space
image plane $F'$. First we can thereby remove the directly
transmitted laser beam by using a beam block located on the
optical axis. As an application of this method of filtering we
consider the reflection of a SPP beam by an in-plane Bragg mirror.
SPPs are launched as before from a gold ridge (50 nm height, 150
nm width) lithographed on a 50 nm thick gold film. The Bragg
mirror \cite{Born} constitutes a one-dimensional lattice of
parallel gold ridges (50 nm height, 140 nm width) separated by a
distance $P$ defining the period of the lattice. The period $P$ is
connected to the SPP wavelength by
$\lambda_{SPP}=2\pi/k_{SPP}<\lambda$ and to the angle of incidence
reflection $\theta_{SPP}$ of the SPP beam relatively to the (in
plane) normal to the lattice by
\begin{equation}
P=\frac{\lambda_{SPP}}{2\cos{\theta_{SPP}}}.
\end{equation}
In the present case shown in Fig.~10 the Bragg mirror is optimized
for $\lambda=800 nm$ (i.~e., $\lambda_{SPP}=785 nm$) and for
$\theta_{SPP}=45^{\circ}$ incidence angle which means
$P\simeq555nm$. The experimental analysis of such a Bragg mirror
when the resonance condition ($\lambda$, $\theta_{SPP}$) is
fulfilled reveals a very high reflectivity of $R\simeq 95$\% (see,
for example, \cite{Harry2} for some earlier results on SPP Bragg
mirrors studied with fluorescence microscopy). However, in the
present experiment we choose an incident angle
$\theta=65^{\circ}$. As a consequence the reflectivity was much
lower (see Fig.~10 A and more details in \cite{Drezet:APL2006}).
\begin{figure}[h]
\includegraphics[width=6cm]{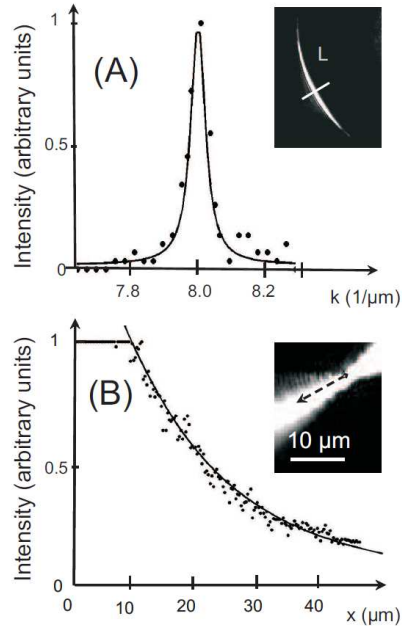}
\caption{(A) Fourier space cross-cut along the short solid line in
Fig.~10 B corresponding to the $L$ beam (see inset). Experimental
data (symbols) are compared with a Lorentz fit (solid curve). The
intensity is normalized by setting the maximum intensity of $L$ to
1. (B) Direct space cross-cut along the L beam in Fig. 10 A. Data
points (symbols) are compared to an exponential fit (solid line).
Data from \cite{Drezet:APL2006} ($\copyright$, Appl.~phys.~Lett.,
American Institute of physics, 2006). }
\end{figure}
This configuration reveals SPP interferences in the vicinity of
the mirror (Fig.~10 A). In Fig.~10 B we show the corresponding
Fourier space image. The different observed arcs of LR rings
correspond to the reflected (R), and transmitted plus incident (T)
beams. The L beam is associated with a SPP launched in the
direction to the left, i.~e, away from the mirror. C is the
directly transmitted laser beam distribution. By acting in the
Fourier plane image $F'$ of the LRM microscope we now block the
information associated with the central beam and with the R or T
beams \cite{Drezet:APL2006}. Thereby the according SPP beam images
are erased from the image plane and consequently weak intensity
beams otherwise observed by interference  can be directly
analyzed. For further analysis we extracted radial cross-cuts
along the white lines as shown in Fig.~10 B \cite{Drezet:APL2006}.
Results are shown in Fig.~11 A for the cross cut along $L$. The
data points agree very well with a Lorentz fit given by Eqs.~20,
21. The FWHM of the Lorentzian distribution gives us a SPP
propagation length of $L_{SPP}=20$ $\mu$m. This value is in
perfect agreement with the cross-cut made along the beam $L$ in
the direct space image 10 A (see Fig.~11 B).
Both data agree also with  values given by the dispersion relations discussed in sect.~1 (see Figs.~2B and 3B).\\
\begin{figure}[h]
\includegraphics[width=8cm]{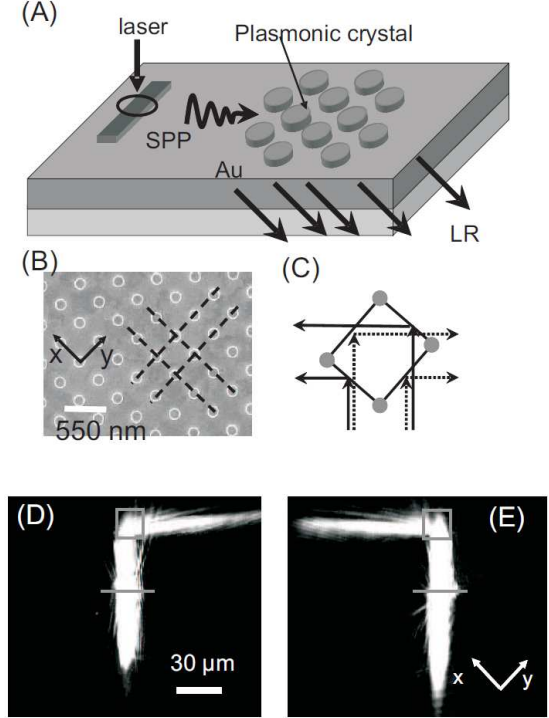}
\caption{(A) Principle of SPP in-plane wavelength demultiplexing
using plasmonic crystal. SPPs launched from a gold ridge (50 nm
height, 150 nm width) are propagating in the direction of a 2D
rectangular lattice (plasmonic crystal) as described in the text.
(B) SEM images of a part of the crystal, Bragg lines are indicated
by dashed lines. Such crystals can be seen as two sets of Bragg
mirrors perpendiculary oriented to each other. (C) Sketch
representing the unit rectangular cell and the direction of the
reflected SPP beam for the two different wavelengths at which the
Bragg reflections occur. (D, E) LRM images of SPP interacting with
the demultiplexer at SPP wavelengths $\lambda_{x}\simeq 730$ nm
and $\lambda_{y}\simeq 784$ nm, respectively. The gold ridge and
plasmonic crystal are indicated by the gray line and the gray box
, respectively. The  vector basis $\mathbf{x}, \mathbf{y}$ of the
latices is indicated in A and E. Data  from
\cite{Drezet:Nanolett2007} ($\copyright$ Nanoletters, American
Chemical Society, 2007). }
\end{figure}
LRM was subsequently applied by us to many SPP in-plane devices
such as beam splitters \cite{Stepanov}, dielectric lenses, prisms
\cite{Hohenau} and wave-guides \cite{Steinberger}. In particular,
LRM experiments were compared to near field optical experiments (
photon scanning tunnelling  microscopy) and showed good agreement
in the cases considered \cite{Hohenau,Steinberger}. LRM appears
thus in this context as a complementary far-field optical method
to NFO such as NSOM. LRM was applied as well for analyzing SPP
Bragg mirrors (with high reflectivity $R\simeq  90-95$\%),
interferometers \cite{Drezet:EPL2006,Drezet:plasmonics2006} and
SPP elliptical cavities \cite{Drezet:APL2005} or 2D SPP
microscopes \cite{submitted}. In this context we observed
\cite{Drezet:EPL2006} stationary SPP waves with very high
visibility
$V=(I_{\textrm{max}}-I_{\textrm{min}})/(I_{\textrm{max}}+I_{\textrm{min}})$
by using LRM. This proves directly that SPP wave coherence is
conserved in LRM and can exploited for quantitative analysis down
to the spatial resolution limit $\lambda_{SPP}/2$. Actually SPP
interferometers such as the ones described in
\cite{Drezet:APL2005,Drezet:plasmonics2006,Harry2} reveal clear
interference pattern and oscillation characteristics of these
devices.
It is thus possible to develop 2D interferometry for SPP waves having all the advantages of current macroscopic interferometry techniques.\\
We also mention the realization of plasmonic crystals (i.~e.,
photonic crystals for SPPs) which were studied using LRM (see
Fig.~12). In such devices \cite{Drezet:Nanolett2007} rectangular
2D latices made of gold  nano-protrusions (200 nm diameter, 50 nm
height) deposited on a  50 nm thick gold film  (see Fig.~12B) are
used to create photonic band gaps at specific SPP wavelengths
$\lambda_{x}\simeq 730$ nm and $\lambda_{y}\simeq 784$ nm  (i.~e.,
laser wavelengths of respectively 750 nm and 800 nm) corresponding
to the two periods of the lattice $P_{x}=\lambda_{x}/\sqrt{2}=
516$ nm and $P_{y}=\lambda_{y}/\sqrt{2}=554$ nm. The existence of
these band gaps implies that SPP plane waves impinging on small
devices build up with such lattice will generate stationary waves
in the crystal. More precisely this implies that SPPs will be
reflected in specific and different directions if their
wavelengths match the values $\lambda_x$ or $\lambda_y$ and if the
angle of incidence relatively to the normal to the Bragg planes of
the crystal (Figs.~12 B, C) equals $45^{\circ}$. Such devices act
consequently as an efficient in-plane wavelength demultiplexer for
SPPs \cite{Drezet:Nanolett2007} as it was indeed observed
experimentally (see Fig.~12D and E). While the results discussed
here were achieved within the visible spectral range, plasmonic
crystal devices are expected to perform even better (e.g., in
terms of spectral selectivity) in the infrared (telecom) spectral
range due to significantly lower ohmic losses
\cite{Nikolajsen:2003}. In general, the use of multiplexers,
splitters and tritters \cite{Drezet:Nanolett2007} in photonic
applications might be specifically appealing due to their small
footprint in the range of $10\times10$ $\mu$m$^2$. Furthermore,
the use as building blocks for classical \cite{Teich} or quantum
\cite{Knill:2001} optical
computing can be envisaged. \\
\section{Conclusion}
In this article we reviewed the field of leakage radiation
microscopy (LRM) theoretically and experimentally. Theoretically
we analyzed how SPP can generate leaky wave in the glass substrate
by tunnelling from the air side through a thin metal film
supporting SPP waves. We showed in  particular that for thick film
($D\geq 70$ nm) leakage radiation (LR) does not affect the
dispersion relation on the air/metal interface. Importantly the
angular distribution of LR is located on a cone matching the SPP
dispersion relation. We also reviewed the first experimental
results reporting the observation of LR  on rough surface and
using near field optics methods. We analyzed more recent
application of LRM to SPP  nano-devices fabricated by electron
beam lithography. From all these results we can conclude that LRM
is a convenient and versatile far field optical method for
analyzing quantitatively SPP propagation on flat film and their
interaction with various nano-devices of direct practical
interest. Such versatility positions LRM as an appealing
alternative to near field optics for studying SPP propagation on a
scale of, or larger than the wavelength.

For financial support the European Union, under project  FP6
2002-IST-1-507879 is acknowledged.

\end{document}